\documentclass[aps,prd,preprint,groupedaddress,showpacs,11pt]{revtex4}

\usepackage{amsmath}
\usepackage{amssymb}
\usepackage{graphics}
\usepackage{graphicx}
\usepackage{psfrag}
\usepackage[mathcal]{eucal}

\def\lsim{\buildrel < \over {_{\sim}}}
\def\gsim{\buildrel > \over {_{\sim}}}

\begin{document}

\preprint{FERMILAB-Pub-06-272-T}
\preprint{UCI-TR-2006-20}


\title{Model Predictions for Neutrino Oscillation Parameters}

\author{Carl H. Albright$^{1,2,}$}
\email{albright@fnal.gov}
\author{Mu-Chun Chen$^{2,3,}$}
\email{mcchen@fnal.gov}
\affiliation{$^1$Department of Physics, Northern Illinois University, DeKalb,
  IL 60115\\
$^2$Fermi National Accelerator Laboratory, Batavia, IL 60510\\
$^3$Department of Physics \& Astronomy,~University of California, Irvine,
  CA 92697}

\date{November 22, 2006}

\begin{abstract}
We have surveyed leptonic and grand unified models of neutrino masses and 
mixings in the literature which are still viable and give numerical 
predictions for the reactor angle, $\theta_{13}$. The results are of 
considerable interest in anticipation of the next generation reactor 
experiments and the possible future need for neutrino factories.  Of the 
63 models considered which were published or posted on the Archive before 
June 2006, half predict values of $\sin^2 2\theta_{13} \gsim 0.015$, which 
should yield positive signals for $\bar{\nu}_e$ disappearance in the 
reactor experiments planned for the near future.  Depending upon the 
outcome of those experiments, half of the models can be eliminated on the 
basis of the presence or absence of such an observed $\bar{\nu}_e$ 
disappearance signal.

\end{abstract}


\maketitle


\section{Introduction}

With the confirmation of atmospheric muon-neutrino oscillations by the 
Super-Kamiokande collaboration~\cite{Fukuda:1998mi}, the need for physics 
beyond the standard model became clear.  This prompted many authors to 
construct mass matrix models to explain the neutrino mixings and the 
still not well-determined neutrino mass spectrum.  Meanwhile underground 
and reactor experiments \cite{solar} confirmed the existence of solar 
electron-neutrino oscillations as a solution to the solar neutrino puzzle, 
whereby the detection of solar electron-type neutrinos was depleted 
\cite{Davis} relative to the standard solar model projections \cite{BP98}.
Since then many of the first round of models have fallen by the wayside, 
as the once-preferred small mixing angle (SMA) solution for the solar 
neutrino oscillations has been replaced by the large mixing angle (LMA) 
solution.  Accelerator experiments have now also contributed to the greatly
increased precision of all observed oscillation results.  In particular,
the observed large atmospheric neutrino mixing has persisted as a nearly 
maximal $\nu_\mu - \nu_\tau$ mixing \cite{maxatm}.  The solar neutrino 
mixing is now known to be large but not maximal \cite{solar2}.  On the other 
hand, only a relatively small upper limit has been placed on the 
$\bar{\nu}_e$ mixing with the other two antineutrinos by two reactor neutrino 
experiments carried out several years ago \cite{reactor}.  Clearly the 
neutrino mixing pattern is totally unlike that observed in the quark sector.

Despite the refinement of the experimental results, many of the more 
recent neutrino models have survived todate due to the uncertainty in the 
precise magnitude for the reactor angle, $\theta_{13}$, and the unknown 
mass hierarchy for the neutrino spectrum: normal, degenerate, or inverted.  
In this paper we are primarily concerned with the model predictions for 
the reactor angle.  While new reactor experiments \cite{newreactor} are being 
planned or are already in construction to measure this angle down to 
$3^\circ$, i.e., $\sin^2 2 \theta_{13} \simeq 0.01$, it is of considerable 
interest to learn whether that reach will be sufficient to determine the 
angle through the detection of a ${\bar \nu}_e$ disappearance signal, 
or whether a neutrino factory will be required which can probe a much 
smaller angle \cite{NuFACT}.  For this purpose, we 
have surveyed 63 models in the literature which are still viable candidates
and have reasonably well-defined predictions for $\theta_{13}$.  
Roughly half of the models predict that $\sin^2 2\theta_{13}$ covers the 
range from 0.015 to the present upper bound of 0.15 \cite{reactor}.
Hence half of the models can be eliminated in the next round of reactor 
experiments, based on the presence or absence of an observed $\bar{\nu}_e$
disappearance signal.  While none of the models proposed so far may be correct,
the distribution of results for $\sin^2 2\theta_{13}$ does provide some 
indication of what one may expect to find.

In Sect. II we define the mixing angles and state the present experimental
results.  A brief description of the types of models based on lepton flavor
symmetries and/or grand unification for the quarks and leptons is presented
in Sect. III.  Several more comprehensive reviews of mass matrix models
exist in the literature \cite{reviews}, and we refer the interested reader 
to them for additional details.  We have restricted our attention to models 
with just three active neutrinos and no sterile neutrinos.  This seems 
justified at present in light of the conflicting evidence for an additional 
oscillation between $\bar{\nu}_\mu$ and $\bar{\nu}_e$ observed by the LSND 
Collaboration \cite{LSND}, but not by the Karmen Collaboration \cite{KARMEN}.  
The MiniBooNE Collaboration is expected to throw more light on this situation 
shortly \cite{MiniBooNE}.  We make clear our acceptance criteria for the 
three flavor neutrino models to be considered and give tables 
of the mixing angles predicted for each in Sect. IV.  
Histograms for $\sin^2 \theta_{13}$ are plotted for all 63 models
together, for the 27 models which give numerical predictions for all three 
mixing angles lying within the 2$\sigma$ experimental bounds, and separately 
for those models with normal or inverted neutrino mass hierarchy.  
Our conclusions are presented in Sect.~V.  

\section{Leptonic Mixing Matrix and Present Experimental Information}

Here for completeness we present a brief description of the well-known 
neutrino mixing matrix and give an up-to-date summary of the constraints
on the mixing angles.  

The light left-handed neutrino flavor states are related to the neutrino 
mass eigenstates by the linear combinations
\begin{equation}
  |\nu_\alpha\rangle = \sum_i (U_{\nu_L})_{\alpha i} |\nu_i\rangle,
  \label{eq:nustates}
\end{equation}

\noindent  where $\alpha = e,\ \mu,\ \tau$ and $i = 1,2,3$.  If neutrinos 
are Majorana particles, the $U_{\nu_L}$ transformation matrix is obtained by 
diagonalization of the effective left-handed Majorana neutrino mass matrix 
according to 
\begin{equation}
  M^{diag}_\nu = U^T_{\nu_L} M_\nu U_{\nu_L},
\label{eq:Majorana}
\end{equation}

\noindent where $M_\nu$ is model-dependent and is typically constructed from 
some basic symmetry principle, or follows
from the seesaw mechanism \cite{seesaw} in grand unified (GUT) models.  
For models in which the light neutrinos are assumed to be Dirac particles, 
the $U_{\nu L}$ unitary transformation matrix can be obtained from the 
bi-unitary transformation which relates the flavor basis to the mass basis 
as in
\begin{equation}
  M^{diag}_\nu = U^\dagger_{\nu_R} M^D_\nu U_{\nu_L}.
\label{eq:Dirac}
\end{equation}

\noindent  The light Dirac neutrino mass matrix, $M^D_\nu$, is also
constructed according to some symmetry principle.  

In models based on some assumed lepton flavor symmetry, the charged lepton 
mass matrix is assumed to be diagonal in the lepton flavor basis with the 
charged lepton masses $m_e,\ m_\mu$ and $m_\tau$ ordered along the diagonal; 
thus $U_{L_L}$, the counterpart of $U_{\nu_L}$, is just the identity matrix.  
For GUT models, on the other 
hand, the charged lepton mass matrix is generally not diagonal in the GUT
flavor basis.  In this case in analogy with Eq. (\ref{eq:Dirac}) above,
$U_{L_L}$ can be determined from the bi-unitary transformation
\begin{equation}
  M^{diag}_L = U^\dagger_{L_R} M_L U_{L_L},
\label{eq:lepton}
\end{equation}

\noindent where we have again adopted the convention that the right-handed
fields act on the left and the left-handed fields on the right of the 
Dirac mass matrix.  For either type of model, the left-handed neutrino 
Pontecorvo-Maki-Nakagawa-Sakata (PMNS) mixing matrix is then given 
by~\cite{PMNS} 
\begin{equation}
  V_{PMNS} \equiv U^\dagger_{L_L}U_{\nu_L} = U_{PMNS}\Phi.
\label{eq:VPMNS}
\end{equation}

\noindent  It is convenient to choose by convention
\begin{equation}
  U_{PMNS} = \left(
  \begin{array}{ccc}
   c_{12}c_{13} & s_{12}c_{13} & s_{13}e^{-i\delta}\\
       -s_{12}c_{23} - c_{12}s_{23}s_{13}e^{i\delta} & 
       c_{12}c_{23} - s_{12}s_{23}s_{13}e^{i\delta} & s_{23}c_{13} \\
       s_{23}s_{12} - c_{12}c_{23}s_{13}e^{i\delta} &
       -c_{12}s_{23} - s_{12}c_{23}s_{13}e^{i\delta} & c_{23}c_{13} 
       \end{array}\right)
\label{eq:UPMNS}
\end{equation}

\noindent in analogy with the quark mixing matrix, along with the 
Majorana phase matrix, 
\begin{equation}
  \Phi = {\rm diag}(e^{i\chi_1},\ e^{i\chi_2},\ 1),
\label{eq:Phi}
\end{equation}

\noindent in terms of the three mixing angles, $\theta_{12},\ \theta_{23}$ and 
$\theta_{13}$; the Dirac $CP$ phase, $\delta$; and the two Majorana phases,
$\chi_1$ and $\chi_2$.  With Dirac neutrinos, one is free to make 
phase transformations on both $U_{L_L}$ and $U_{\nu_L}$, so $\Phi$ is just
the identity matrix.  But in the Majorana neutrino case where $U_{\nu_L}$
is defined by Eq. (\ref{eq:Majorana}), an arbitrary phase transformation is 
not possible when one demands real diagonal neutrino mass entries.  Hence
the presence of the Majorana phase matrix is required in order to adopt the
convention for $U_{PMNS}$ specified in Eq. (\ref{eq:UPMNS}).

We now summarize the numerical information for these mixing parameters, as 
given by Maltoni, Schwetz, Tortola, and Valle~\cite{Maltoni:2004ei} in a recent
updated global analysis which incorporates all the latest results 
cited in \cite{Davis}-\cite{reactor}.  Within 2$\sigma$ accuracy, they found
\begin{equation}
\begin{array}{rcl}
    \Delta m^2_{21}&=&(7.3-8.5)\times 10^{-5}\ {\rm eV^2},\\
    \Delta m^2_{31}&=&(2.2-3.0)\times 10^{-3}\ {\rm eV^2},\\
    \sin^2 \theta_{12}&=&0.26 - 0.36,\\
    \sin^2 \theta_{23}&=&0.38 - 0.63,\\
    \sin^2 \theta_{13}&\leq&0.025.\\
\end{array}
\label{eq:data}
\end{equation}

\noindent No information exists for the Dirac or Majorana $CP$ phases, for
the neutrino mass hierarchy, or for the Dirac {\it vs.} Majorana nature of the 
neutrinos.  Of special interest to us is the upper bound on 
$\sin^2 \theta_{13}$ which has mainly been determined by a non-observation 
of a depletion of the $\bar{\nu}_e$ flux from the CHOOZ reactor \cite{reactor}.
In Sect. IV we shall turn to the model selection criterion we have used to 
extract the various model predictions.  But first we give descriptions of the 
types of models that have been proposed.

\section{Descriptions of Models}

Most neutrino mass matrix models fall into two broad classes: those based on
a lepton flavor symmetry which applies only to the charged lepton and light 
neutrino mass matrices and those based on some grand unification scheme 
which applies to both leptons and quarks.  The latter class typically involves
some family unification group which unites the quarks and leptons at the 
high gauge unification scale, while a flavor symmetry applying to the 
corresponding members of different families may or may not be invoked.  
Moreover, right-handed singlet neutrinos belong to irreducible 
representations and provide an exquisite way to give ultralight masses via 
the seesaw mechanism \cite{seesaw} to the observed left-handed neutrinos.  
In the purely leptonic models some lepton flavor symmetry is generally 
considered, but the ultralight $0.001 - 0.1$ eV neutrino mass scales 
typically remain a puzzle.  Several symmetry schemes have been proposed 
within each of the two classes.  Exceptional cases involve a model based on 
anarchy with no flavor symmetry and models with sequential right-handed 
neutrino dominance but no unification group identified.

Of considerable interest is whether a given model exhibits a normal or 
inverted mass hierarchy.  Models of either hierarchy can be found in several 
categories, while for others only one or the other is definitely preferred.
We shall give broad and general descriptions here for each type of model 
considered.   As cited earlier, we refer the reader to the comprehensive
reviews \cite{reviews} for more details.

\subsection{Anarchy -- a Model of Flavor with No Flavor Symmetry}

Contrary to our general study of mixing predictions of neutrino mass matrix 
models with some specified symmetry, we first consider a model
based on neutrino mass anarchy, where the neutrino mass matrix is completely
random as no flavor symmetry is specified \cite{anarchy}.  With this assumption
of no fundamental distinction among the three flavors of neutrinos and hence 
no preferred basis for the neutrino states, one can simply examine 
statistically the neutrino mixing matrix.  Initially the authors of 
\cite{anarchy} first used a Monte Carlo analysis to test how many sets of 
mixing angles passed imposed cuts and then later applied the invariant Haar
measure to the mixing angles.  Still later de~Gouvea and Murayama \cite{dGM} 
refined the analysis by applying the Kolmogorov-Smirnov statistical test to 
the single $\sin^2 \theta_{13}$ variable to obtain a more precise lower bound 
on its expected value which is quoted in the first table. 

\subsection{Models with Lepton Flavor Symmetries}

It is customary for models of this type to be formulated in the 
leptonic flavor basis for which the charged lepton mass matrix is diagonal.
The differentiating feature of these models then resides solely in the light
left-handed neutrino mass matrix.

\subsubsection{$\mu - \tau$ Interchange Symmetry and $L_e - L_\mu - L_\tau$ 
Conserved Flavor Symmetry}

The most general neutrino mass matrix exhibiting a $\mu-\tau$  
symmetry is given by,
\begin{equation}
M_{\nu} = \left(\begin{array}{ccc}
a & b & b
\\
b & c & d
\\
b & d & c
\end{array}\right) \; .
\end{equation}
In the restricted case where $a, \, b < c \simeq d$, with $M_{\nu}$ having 
the texture
\begin{equation}
M_{\nu} = \frac{\sqrt{\Delta m_{atm}^{2}}}{2} \left(\begin{array}{ccc}
a^{\prime} \epsilon & b^{\prime} \epsilon & b^{\prime} \epsilon
\\
b^{\prime} \epsilon & 1 + \epsilon & -1
\\
b^{\prime} \epsilon & -1 & 1+\epsilon
\end{array}\right) \; ,
\end{equation}
a normal mass hierarchy is obtained when $a^{\prime}$ and $b^{\prime}$ are of 
order one and the size of $\epsilon$ is determined by the ratio, 
$\sqrt{\Delta m_{sol}^{2}/\Delta m_{atm}^{2}}$, multiplied by an order 
one coefficient which is a function of $a^{\prime}$ and $b^{\prime}$. 
One finds $\Delta m_{32}^{2} \simeq m_{3}^{2}$, $\Delta m_{21}^{2} = m_{2}^ 
{2} - m_{1}^{2} > 0$, $\sin^{2} 2\theta_{23} = 1$, $\tan^2\theta_{12}  
\simeq \frac{2\sqrt{2}b^{\prime}}{1-a^{\prime}}$, and $\sin\theta_{13} = 0$.

On the other hand, for the simple form $a = c = d = 0$, the symmetry 
remains unbroken and an inverted mass hierarchy is obtained.  
In this case $\Delta m^2_{32} = \Delta m^2_{31} = m^2_1,\ 
\Delta m^2_{12} = 0$, and $\sin^2 2\theta_{23} = \sin^2 2\theta_{12} = 1$ 
with $\sin^2 2\theta_{13} = 0$, as both the atmospheric and solar neutrino 
mixings are maximal, while the reactor neutrino mixing vanishes.  Neither of 
these sets of predictions are observed experimentally, so soft symmetry
breaking of either matrix texture must be introduced to obtain an acceptable
model with the initial hierarchy unaltered.

The special inverted hierarchy texture case cited above actually exhibits
an enhanced $L_e - L_\mu - L_\tau$ symmetry which is more generally of the 
form~\cite{Mohapatra:2004mf}
\begin{equation}
  M_\nu = \left(
  \begin{array}{ccc} 0 & b & b' \\
   b & 0 & 0 \\
    b' & 0 & 0 \end{array}\right).
\label{e-mu-tau}
\end{equation}

\noindent Independent of the relative magnitudes of $b$ and $b'$, this 
rank-2 matrix leads to a neutrino mass hierarchy which is inverted.  Again 
soft symmetry breaking must be introduced in order for the model to be 
experimentally viable.  Examples of these two lepton flavor symmetries are 
grouped together in the tables presented.

\subsubsection{$S_3$ Lepton Flavor Symmetry}

In the case of $S_3$ lepton flavor symmetry involving the permutation group
of three flavors applied to both rows and columns of the neutrino mass matrix,
the most general texture is~\cite{Harrison:1994ww}
\begin{equation}
  M_\nu = \left(\begin{array}{ccc} 
  a & b & b \\
   b & a & b \\
    b & b & a \end{array}\right),
\label{S3}
\end{equation}

\noindent in terms of two independent parameters, $a$ and $b$. 
Alternatively, one can consider as a basis in the flavor space the unit mass 
matrix with $1$'s down the diagonal and the democratic mass matrix with all 
unit elements.  For the unit matrix, the mass spectrum is clearly degenerate
and all mixing angles vanish.  On the other hand, the rank-1 democratic 
matrix yields a normal mass hierarchy.  Introduction of soft $S_3$-breaking
terms involving the parameter $c$ as in  
\begin{equation}
  M_\nu = \left(\begin{array}{ccc}
    a & b & b \\
     b & a-c & b+c \\
      b & b+c & a-c \end{array}\right)
\label{S3-c}
\end{equation}

\noindent still respects the $\mu - \tau$ exchange symmetry and yields a 
neutrino mixing matrix which has the tribimaximal form suggested by Harrison, 
Perkins, and Scott~\cite{Harrison:2002er}:
\begin{equation}
\label{tribimax}
  U_{PMNS} = \left(\begin{array}{ccc}
  2/\surd{6} & 1/\surd{3} & 0 \\
             -1/\surd{6} & 1/\surd{3} & 1/\surd{2} \\
	   -1/\surd{6} & 1/\surd{3} & -1/\surd{2}\end{array}\right),
\end{equation}

\noindent corresponding to the mixing relations $\sin^2 2\theta_{23} = 1,
\ \sin^2 \theta_{12} = 1/3$, and $\sin^2 \theta_{13} = 0$.  These 
results are close to their experimental values, so only small corrections
may be required.  However, the diagonal charged lepton mass matrix clearly
does not obey the $S_3$ symmetry being considered.

\subsubsection{$A_4$ Lepton Flavor Symmetry}

The permutation group $S_4$ of four objects was first considered by Ma
and Rajasekaran \cite{A4}
%
%
as a discrete flavor symmetry. Its non-Abelian subgroups are 
$S_3, D_4$ and $A_4$.  What makes the subgroup $A_4$ of the twelve even 
permutations of $S_4$ of particular interest is the fact that it is also the 
smallest discrete subgroup of $SO(3)$ which has at least one three-dimensional 
representation.  In fact, there are just four irreducible representations, 
one triplet and three singlets.  The three lepton doublets can be placed
in the ${\bf 3}$ while the three right-handed charged leptons are each 
placed in one of the singlets of $A_4$.  With three Higgs doublets also
transforming as a triplet, one can construct the charged lepton mass matrix.
On the other hand, if the right-handed neutrinos are placed in a triplet
representation, the Dirac and Majorana neutrino mass matrices can be  
generated with Higgs singlets.  This application of $A_4$ and other 
variations show that tribimaximal mixing of neutrinos 
can also be achieved while alleviating some of the problems of the $S_3$ 
flavor symmetry.

\subsubsection{Other Lepton Flavor Symmetries}

Other lepton flavor symmetries have also been considered in the literature
including, for example, $SO(3)$ and $SU(3)$.  Another very popular starting
point involves the arbitrary assignment of texture zeros in the light neutrino 
mass matrix \cite{zeros}.  By doing so, one eliminates those assignments 
which do not yield a neutrino mixing matrix mimicing the nearly tribimaximal 
mixing form.  The positions of the texture zeros may then point the way to some
underlying flavor symmetry.  Some of these models are included in our study.

Various attempts have also been made to extend the lepton flavor symmetries 
proposed to the quark sector.  But without a grand unification symmetry 
framework, they have met with rather mixed results.  

\subsection{Sequential Right-Handed Neutrino Dominance}

Another class of models which do not neatly fit into the lepton flavor 
symmetry class or the grand unification class are models with right-handed
neutrino dominance~\cite{seqdom}.  Here three right-handed neutrinos are 
introduced which have a strong hierarchical mass spectrum.  In the absence 
of any family gauge symmetry, one can still deduce that the light neutrino mass
spectrum is controlled by the sequential dominance of the right-handed 
neutrinos, i.e., the mass of the heaviest left-handed neutrino is determined
largely by the mass of the lightest right-handed neutrino, etc.  With this
type of model, authors have shown that near tribimaximal mixings
can also be obtained.

\subsection{Grand Unified Models}

An alternative approach is to start with a vertical family unification
symmetry at some grand unification scale~\cite{Chen:2003zv}.  One can then try 
to impose a
flavor symmetry on this structure which relates the corresponding members
of each family; however, in many cases, one simply adopts an effective operator
approach which arbitrarily assigns certain operators to each element of 
the mass matrices.  Models formulated in this framework are obviously 
much more ambitious than the previous ones dealing only with the lepton
sector. The mass and mixing results for the leptons are highly constrained
by the input parameters introduced for the quark sector, since the Yukawa
couplings apply to both quark and lepton Dirac mass matrices.  Nevertheless,
successful models exist in the literature which are still viable, even after 
tighter constraints on both the quark and lepton mixing parameters have 
been obtained experimentally.  We briefly discuss models in the following 
categories.

\subsubsection{$SU(5)$ and Flipped $SU(5)$ Models}

Grand unification with a high scale $SU(5)$ symmetry was first proposed by 
Georgi and Glashow \cite{gg}, where the quarks, leptons and left-handed 
neutrinos can be placed into ${\bf 10}$ and $\overline{\bf 5}$ representations.
With the appearance of neutrino oscillations it was then suggested to 
place the right-handed neutrinos into $SU(5)$ singlets.   The Higgs fields 
are conventionally placed in the $\overline{\bf 5}$ and ${\bf 24}$
representations.  This group symmetry with a minimal Higgs structure proved 
to be less interesting when the limit on the proton decay lifetime increased 
several orders of magnitude above the predicted range of $10^{29} - 10^{30}$ 
years. 

An alternative procedure is to consider flipped $SU(5)$ \cite{flipped} in 
which the charged lepton and right-handed neutrino are interchanged in the 
${\bf 10}$ and ${\bf 1}$ representations, as well as the conjugate up and 
conjugate down quarks in the ${\bf 10}$ and $\overline{\bf 5}$ representations 
with respect to the usual $SU(5)$ assignments.  While models of this type have 
been pursued in the literature, none of them have numerical predictions 
for the reactor angle $\theta_{13}$, so we do not elaborate on them further.

\subsubsection{$SO(10)$ Models with Higgs in High Rank Representations}

The $SO(10)$ grand unification symmetry is an economical and attractive one
\cite{Chen:2003zv}, for all sixteen left-handed quark and lepton fields and 
their left-handed conjugates fit neatly into one ${\bf 16}$ representation per 
family.  Many models exist in the literature which differ from one another by 
their Higgs representation assignments and flavor symmetry imposed, if any.  
To appreciate this, it is of interest to note the following decompositions of 
the direct product of representations:
\begin{equation}
\begin{array}{rcl}
  {\bf 16 \otimes 16} &=& {\bf 10_s \oplus 120_a \oplus 126_s}, \\
  {\bf 16 \otimes \overline{16}} &=& {\bf 1 \oplus 45 \oplus 210}, \\
\end{array}
\label{products}
\end{equation}

\noindent where in the first product the ${\bf 10}$ and ${\bf 126}$ matrices
are symmetric, while the ${\bf 120}$ is antisymmetric.

For the simplest Higgs structure \cite{babumohapatra} one can assume the 
presence of one ${\bf 10}_H$, one $\overline{\bf 126}_H$, and one 
${\bf 126}_H$ representations, where the latter is needed to preserve 
a D-flat direction at the GUT scale in supersymmetric models.  
The ${\bf 10}_H$ contains two Higgs doublets, which appear in the ${\bf 5} +
\overline{\bf 5}$ $SU(5)$ decomposition, and contribute to the Dirac mass 
matrices, while the $\overline{\bf 126}_H$ contains a Higgs singlet which
contributes to the right-handed Majorana neutrino mass matrix.  Recently 
additional Higgs fields in the ${\bf 120}_H$, ${\bf 210}_H$, and ${\bf 45}_H$
representations have also been considered by model builders in order to 
fine tune their predictions.  The Dirac mass matrices will then have 
symmetric or antisymmetric textures, or even matrix elements with linear 
combinations of the two forms.  The high ranks of these Higgs representations,
rank-3, 4, and 5 are somewhat disfavored in the string theory framework
\cite{strings}.

With the vacuum expectation value of the $\overline{\bf 126}_H$ appearing near
the GUT scale and giving massive entries to the right-handed Majorana mass 
matrix, the conventional type I seesaw 
mechanism~\cite{seesaw} provides a ready explanation
for the ultralight left-handed neutrinos:
\begin{equation}
\label{typeI}
    M_\nu = - M^T_N M^{-1}_{R} M_N,
\end{equation}

\noindent where $M_N$ is the Dirac neutrino mass matrix and $M_{R}$ is the 
right-handed Majorana matrix in the basis convention that the conjugate 
left-handed fields appear on the left and the left-handed fields on the 
right of $M_N$.  It is easy to show that models involving a type I seesaw
yield a normal mass hierarchy for the light neutrinos \cite{normal}.  

Some authors also allow the possibility that Higgs triplet VEV's exist in the 
$\overline{\bf 126}_H$ and can give non-zero entries to the left-handed 
Majorana mass matrix, $M_{LL}$.  One then generates the light 
left-handed neutrinos through the type II (or mixed) seesaw 
mechanism~\cite{typeII}:
\begin{equation}
\label{typeII}
    M_\nu = M_{LL} - M^T_N M^{-1}_R M_N.
\end{equation}

\noindent  In this case, the light neutrino mass hierarchy can be normal, 
degenerate, or  inverted, the latter two occuring when the $M_{LL}$ 
contributions are comparable to or larger than the type I contributions.

One possible breaking pattern of SO(10) down to the SM gauge group 
has the Pati-Salam group~\cite{p-s}, $SU(4)\times SU(2)_{L} \times 
SU(2)_{R}$, as the intermediate gauge group,
\begin{equation}
SO(10) \rightarrow SU(4)\times SU(2)_{L} \times SU(2)_{R} 
\rightarrow SU(3) \times SU(2)_{L} \times U(1)_{Y}
\end{equation}
This breaking can be achieved with a minimal Higgs content that has  
one ${\bf 10}_H$, one ${\bf 45}_H$, one ${\bf 54}_H$ and a conjugate pair of 
${\bf 126}_H \oplus \overline{\bf 126}_H$. Due to the left-right symmetry, 
the resulting mass matrices are symmetric. In addition, if the minimal  
Higgs content described above is utilized, one has the following  
relations: the up type quark mass matrix and the Dirac neutrino mass  
matrix are identical, while the mass matrix of the down type quarks  
and that of the charged leptons are identical, up to some calculable  
Clebsch-Gordan coefficients which, when combined with the family symmetry, 
can be used to obtain the Georgi-Jarlskog relations \cite{g-j}
required by phenomenology. The Majorana mass terms for the right-handed  
neutrinos arise from coupling to the $\overline{\bf 126}_H$. These 
intra-family relations among the mass matrices greatly reduce the number of  
free parameters in the Yukawa sector, making these models very  
predictive.

\subsubsection{$SO(10)$ Models with Lopsided Mass Matrices}

While the minimal Higgs models discussed above naturally preserve R-parity
when the $\overline{\bf 126}_H$ develops a VEV and lepton number is violated
by two units, they suffer from the disadvantage that they become 
non-perturbative above the GUT scale due to the high rank of the 
representations.  Models which do not share this problem can be constructed 
by using lower rank Higgs representations ~\cite{lopsided} including: 
${\bf 10}_H$'s, one ${\bf 45}_H$, and one or two pairs of ${\bf 16}_H$ - 
$\overline{\bf 16}_H$.  At the GUT scale,
VEV's of the ${\bf 45}$ and the $SU(5)$ singlet parts of a ${\bf 16}_H$ - 
$\overline{\bf 16}_H$ pair break the $SO(10)$ symmetry to that of the 
standard model.  Near GUT scale masses are generated for the right-handed 
neutrinos by pairs of $\overline{\bf 16}_H$ $SU(5)$-singlet VeV's which form 
an effective $\overline{\bf 126}_H$.  Due to the nature of these VEV's, 
lepton number is broken but only by one unit, so R-parity is broken.  
Hence it is necessary to introduce a matter parity in order to preserve the 
distinction between particles and their super-partners, unlike in the 
higher-rank Higgs type of models described above where R-parity is 
automatically preserved.

Vacuum expectation values for the doublets in the $5({\bf 10}_H)$, 
$\bar{5}({\bf 10}_H)$, and $\bar{5}({\bf 16}_H)$ are then assumed to be 
generated at the electroweak scale.  Since the surviving $\bar{5}$ $v_d$ 
VEV is a linear combination of the VEVs of 
the ${\bf 10}_H$ and ${\bf 16}_H$, one finds that $\tan \beta = v_u/v_d$ can 
be in the range of 5 - 55 rather than simply 55 when the ratio involves just 
the two doublets from the ${\bf 10}_H$.  While the two doublets in the 
$5({\bf 10}_H)$ and $\bar{5}({\bf 10}_H)$ Higgs representations contribute 
in a symmetric way to the $ij$ components 
of the Dirac neutrino and up quark mass matrices, and to the charged lepton
and down quark mass matrices, respectively, the doublet in the 
$\bar{5}({\bf 16}_H)$ representation contributes only to the charged lepton
and down quark mass matrices in a lopsided fashion.  This follows because
$d_L$ and $\ell^c_L$ lie in a $10({\bf 16})$, while $d^c_L$ and $\ell_L$ lie 
in a $\bar{5}({\bf 16})$ matter representation.  The complete 
Froggatt-Nielsen tree diagram then makes clear that if a charged lepton 
$ij$ mass matrix element receives a large contribution while the transposed
element $ji$ vanishes, the opposite will be true for the down quark mass 
matrix.  This lopsided behavior for the charged lepton mass matrix can lead to 
a large lepton flavor violation in $\mu \rightarrow e + \gamma$, for 
example.  The corresponding branching ratio for the higher rank Higgs models 
tends to be one or two orders of magnitude smaller.

In order to obtain a successful $SO(10)$ GUT model of either type, one must 
not only be able to generate appropriate neutrino masses and mixings, but 
the quark masses and CKM mixings for the quark sector must agree with the 
observed values after evolution downward from the GUT scale.  This imposes 
considerably more constraints on the model than are present with the purely 
leptonic models discussed in part A.  For either type of $SO(10)$ model, 
the appearance of a PMNS neutrino mixing matrix close to the tribimaximal 
mixing form is usually regarded as accidental, rather than reflecting a 
symmetry inserted at the outset as in the purely leptonic flavor models.

\subsubsection{$E_6$ and $E_8 \otimes E_8$ Models}

Some authors have pursued models based on the exceptional $E_6$ gauge group.
The matter fields of interest are placed in the ${\bf 27}$ dimensional 
representation, while the Higgs fields are placed in a ${\bf 27}_H$, 
${\bf 351}_H$, and/or ${\bf 351'}_H$ representations.  One is then faced 
with the problem of making massive many of the extra fields which are 
present in such high dimensional representations.  Some progress has been 
made, but none of the models have any firm numerical predictions for the 
neutrino mixing angles, so we do not consider them further.

Even more ambitious models have attempted to deal with $E_8 \otimes E_8$
grand unified models which naturally arise in the heterotic string theory.
One of the $E_8$'s is assumed to break down to $E_6 \times SU(3)$, while 
the other represents a hidden symmetry.  Many of these models, as well as
the $E_6$ models discussed in the previous paragraph, are formulated in 
five or six dimensions. Again no firm numerical predictions for the mixing
angles have been obtained. 

\section{Results for the Models Surveyed}

In the previous Sect. we have presented broad general descriptions of models 
in the categories considered todate.  Here we present results for the
various models in the literature.  We begin by defining our selection criteria 
for the three neutrino flavor models to be included in our survey.  
First of all we require that 
the models give the LMA solution for the solar neutrino oscillations and that 
firm and reasonably restrictive numerical predictions be given for the reactor 
$\sin^2 \theta_{13}$ mixing parameter.  We do not require that the other 
two mixing angles or the mass squared differences be predicted, but all 
mixing angles for which information is given are listed in the tables.
Models which are clearly in conflict with the present neutrino oscillation
data are not considered, though we have not imposed an upper limit on the 
prediction of $\sin^2 \theta_{13}$.  Many of the models have evolved with 
time and have been updated by their authors.  As such we have generally 
listed only the latest published or archived version, except in cases where 
some important variation has provided two noticeably different results for 
the reactor angle.  Only single references are given in alphabetical order 
to the accepted models in each category.  The 
interested reader can readily track down earlier references to each model,
if they exist, and can use them to learn the specific details of a certain
model.   Finally, we note that we have arbitrarily selected May 2006 as the 
cutoff date in accepting models for our compilations.

In Table I we list 26 models cited in \cite{dGM} and \cite{BM} - \cite{WY} 
which exhibit one of the lepton flavor symmetry types: anarchy, 
$L_e - L_\mu-L_\tau$, $S_3$ or $S_4$, $A_4$, $SO(3)$, and texture zeroes.
Only six of them have firm predictions for all three mixing angles.  The 
predictions for $\sin^2 \theta_{13}$ cover the full range of possibilities,
from the present upper bound of 0.025 for the CHOOZ limit down to $10^{-5}$
or less which would clearly require information from a neutrino factory to 
measure the result accurately.  

In Table II we list seven models cited in \cite{D} - \cite{K} based on 
sequential right-handed neutrino dominance in which no particular GUT model 
is assumed.  All of these models have a normal hierarchy with a restricted 
range of predictions for the reactor angle, i.e., $\sin^2 \theta_{13} 
\gsim 10^{-3}$.

In Table III we list 24 $SO(10)$ models cited in \cite{BaMa} - \cite{YW} 
based on Higgs fields in the ${\bf 10}_H$, $\overline{\bf 126}_H$, 
${\bf 126}_H$, and possibly ${\bf 120}_H$ or ${\bf 45}_H$
dimensional representations.  As such, their mass matrix elements receive
symmetric or antisymmetric contributions.  Note that all models based on a 
type~I seesaw mechanism have normal hierarchy, whereas those with inverted
hierarchy would be highly unstable \cite{normal}.  Only the type~II seesaw 
models permit a stable inverted hierarchy depending upon the interplay of 
the type I seesaw and left-handed Majorana $M_{LL}$ contributions.  Only three 
of the models predict a value of $\sin^2 \theta_{13} \lsim 10^{-3}$.  

Finally in Table IV are listed six $SO(10)$ models \cite{A} - \cite{P} 
involving Higgs fields in the ${\bf 10}_H$, ${\bf 16}_H$, 
$\overline{\bf 16}_H$, and ${\bf 45}_H$ dimensional representations which 
exhibit lopsided entries for the down quark and charged lepton mass matrices.  
Four of them predict values for $\sin^2 \theta_{13} \lsim 3 \times 10^{-3}$.  
The other two have predictions so near to the CHOOZ bound, they are on 
the verge of being ruled out. 

In order to illustrate these results better visually, we have plotted 
histograms for the $\sin^2 \theta_{13}$ predictions.  In Fig. 1 we show
the results for all 63 models.  Only the results for the 27 models which 
predict all three mixing angles and which lie within the 2$\sigma$ bounds of
Eq. (8) are plotted in Fig. 2.  In Figs. 3a and b
we separate the models into those with normal and inverted hierarchies,
respectively.  On the $\log(\sin^2 \theta_{13})$ scale, we have divided
each power of ten interval into three equal parts.  For those models which
give a range of values which occupy several intervals, we have rescaled
them, so that each model has the same normalized area on each histogram
for the predicted hierarchy.

Clearly the majority of models prefer normal hierarchy, with the two main
exceptions being models based on $L_e - L_\mu - L_\tau$ symmetry or 
those with well-designed texture zeros.  Roughly half of the 63 models 
have $\sin^2 \theta_{13} \gsim 0.004$, or $\sin^2 2\theta_{13} \gsim 0.015$.
This feature remains true even for the smaller number of 27 models which
can accurately predict all three mixing angles.  Since the next generation 
of reactor experiments is expected to reach values of order 0.01 for 
$\sin^2 2\theta_{13}$, we can expect that roughly half of the models will
be eliminated based on the presence or absence of an observed $\bar{\nu}_e$ 
disappearance signal.  The possible need for a neutrino factory to reach 
even smaller values of the reactor neutrino mixing angle, if necessary,
will then become apparent.  On the other hand, if a disappearance signal 
is seen, and its value for $\sin^2 \theta_{13}$ can be well measured, the 
number of surviving models will be greatly reduced.  Determination of the mass 
hierarchy will narrow the number down even further.

\section{Conclusion}

From our survey we found that the predictions for the angle $\theta_{13}$ 
range from zero to the current experimental upper limit.  For models based on 
GUT symmetries, normal mass hierarchy can be generated naturally. Inverted 
hierarchy may also be obtained in these models with a type-II seesaw, even 
though some fine-tuning is needed. Predictions for the mixing angle 
$\theta_{13}$ in these models tend to be relatively large, with a median value 
$\sin^{2}2\theta_{13} \simeq 0.015$. On the other hand, models based on 
leptonic symmetries can give rise to inverted mass hierarchy, and the 
predictions for $\theta_{13}$ can be quite small. Therefore, if the inverted 
mass hierarchy is observed experimentally and the mixing angle $\theta_{13}$  
turns out to be tiny, this experimental evidence will then give strong 
support to models based on lepton symmetries.  However, if $\theta_{13}$ turns 
out to be relatively large, one will not be able to tell the two different 
classes apart. A precise measurement for the deviation of $\theta_{23}$ from 
$\pi/4$ can also be crucial for distinguishing different models.  This is 
especially true for models based on lepton symmetries in which the deviation 
strongly depends on how the symmetry breaking is introduced into the models.  
Clearly precision measurements are indispensable in order to distinguish 
different classes of models and to narrow down the number of acceptable 
models.\\[0.2in]

\begin{center}
  {\bf ACKNOWLEDGMENTS}
\end{center}

The authors thank Stephen Geer for encouraging them to pursue this work and
for his helpful suggestions during the course of the work.  Fermilab is 
operated by Universities Research Association Inc.under contract No. 
DE-AC02-76CH03000 with the Department of Energy.


\begin{table}
\caption{\label{TableI} Mixing Angles for Models with Lepton Flavor Symmetry.}
\vspace*{0.1in}
\begin{tabular}{lr|c|c|c|c}
\multicolumn{2}{l|}{\large Reference} & {\large Hierarchy}  & 
  {\large \ $\sin^2 2\theta_{23}$ \ } & {\large \ $\tan^2 
  \theta_{12}$ \ } & {\large $\sin^2 \theta_{13}$}\\[0.1in]
\hline \hline 
\multicolumn{3}{l|}{\bf Anarchy Model:} & & \\[-0.05in]
dGM & \cite{dGM}	& Either	&	&	& $\geq 0.011\ @\ 2\sigma$
  \\[0.05in]
\hline
\multicolumn{3}{l|}{\bf $\mathbf{ L_e - L_\mu - L_\tau}$ Models:} & & 
  \\[-0.05in]
BM & \cite{BM} 	& Inverted & 	&	& 0.00029\\[-0.05in]
BCM & \cite{BCM}	& Inverted &	&	& 0.00063\\[-0.05in]
GMN1 & \cite{GMN1}	& Inverted &	& $\geq 0.52$ & $\leq 0.01$ \\[-0.05in]
GL & \cite{GL}	& Inverted & 	&	& 0 \\[-0.05in]
PR & \cite{PR}	& Inverted & 	& $\leq 0.58$	& $\geq 0.007$ 
  \\[0.05in]
\hline
\multicolumn{3}{l|}{\bf $\mathbf{S_3}$ and $\mathbf{S_4}$ Models:} & & 
  \\[-0.05in]
CFM & \cite{CFM}	& Normal	&	&	& \ 0.00006 - 0.001 
  \\[-0.05in]
HLM & \cite{HLM}	& Normal	& 1.0	& 0.43	& 0.0044 \\[-0.05in]
      &    & Normal  & 1.0	& 0.44	& 0.0034 \\[-0.05in]
KMM & \cite{KMM}	& Inverted & 1.0	&	& 0.000012 \\[-0.05in]
MN & \cite{MN}	& Normal	& 	&	& 0.0024 \\[-0.05in]
MNY & \cite{MNY}	& Normal	&	&	& 0.000004 - 0.000036 
  \\[-0.05in]
MPR & \cite{MPR}	& Normal  & 	&	& 0.006 - 0.01 \\[-0.05in]
RS & \cite{RS}	& Inverted & $\theta_{23} \geq 45^\circ$ &  & $\leq 0.02$ 
  \\[-0.05in]
     &     & Normal	& $\theta_{23} \leq 45^\circ$ & & 0 \\[-0.05in]
TY & \cite{TY}	& Inverted & 0.93	& 0.43 	& 0.0025 \\[-0.05in]
T & \cite{T}	& Normal	& 	&	& \ 0.0016 - 0.0036 
  \\[0.05in]
\hline
\multicolumn{3}{l|}{\bf $\mathbf{A_4}$ Tetrahedral Models:} & &  \\[-0.05in]
ABGMP & \cite{ABGMP}& Normal	& 0.997 - 1.0  & 0.365 - 0.438  & 
  0.00069 - 0.0037 \\[-0.05in]
AKKL & \cite{AKKL}	& Normal	&	&	& 0.006 - 0.04 \\[-0.05in]
Ma & \cite{Ma}	& Normal	& 1.0	& 0.45	& 0 \\[0.05in]
\hline
\multicolumn{3}{l|}{\bf $\mathbf{SO(3)}$ Models:} & & \\[-0.05in]
M & \cite{Mas}	& Normal	& 0.87 - 1.0 & 0.46 & 0.00005 \\[0.05in]
\hline
\multicolumn{3}{l|}{\bf Texture Zero Models:} & & \\[-0.05in]
CPP & \cite{CPP}	& Normal	& 	&	& 0.007 - 0.008 \\[-0.05in]
       &   & Inverted &	&	& $\geq 0.00005$ \\[-0.05in]
       &   & Inverted &	&	& $\geq 0.032$ \\[-0.05in]
WY & \cite{WY}	& Either	&	&	& 0.0006 - 0.003 \\[-0.05in]
       &   & Either  &	&	& 0.002 - 0.02 \\[-0.05in]
       &   & Either  &	&	& 0.02 - 0.15 \\[0.05in]
\hline \\[0.2in]
\end{tabular}
\end{table}

\begin{table}
\caption{\label{TableII} Mixing Angles for Models with Sequential Right-Handed 
  Neutrino Dominance.}
\vspace*{0.1in}
\begin{tabular}{lr|c|c|c|c}
\multicolumn{2}{l|}{\large Reference\ } & {\large \  Hierarchy \ } & 
  {\large \  $\sin^2 2\theta_{23}$ \ } & {\large \  $\tan^2 
  \theta_{12}$ \ } & {\large \  $\sin^2 \theta_{13}$ \ }\\[0.1in]
\hline \hline
D & \cite{D}	& Normal	& 	&	& 0.008 - 0.14 \\[-0.05in]
EH & \cite{EH}	& Normal	& 0.98	& 0.32	& 0.014 \\[-0.05in]
     &     & Normal	& 0.98	& 0.34	& 0.012 \\[-0.05in]
     &     & Normal	& 0.99	& 0.45	& 0.0009 \\[-0.05in]
     &     & Normal	& 0.97 	& 0.30	& 0.014 \\[-0.05in]
H  & \cite{H}	& Normal	& 1.0	& 0.42	& 0.0033 \\[-0.05in]
K & \cite{K}	& Normal	& 0.99 - 1.0	& 0.40 - 0.62 & 0.0027 \\
\hline 
\end{tabular}
\end{table}

\vspace*{-0.5in}
\begin{table}
\caption{\label{TableIII} Mixing Angles for $SO(10)$ Models 
  with Symmetric/Antisymmetric Contributions.} 
\vspace*{0.1in}
\begin{tabular}{lr|c|c|c|c}
\multicolumn{2}{l}{\large Reference\ } & {\large \  Hierarchy \ } & 
  {\large \  $\sin^2 2\theta_{23}$ \ } & {\large \  $\tan^2 
  \theta_{12}$ \ } & {\large \  $\sin^2 \theta_{13}$ \ }\\[0.1in]
\hline \hline
BaMa & \cite{BaMa}	& Normal	& 0.88	& 0.33	& 0.015 - 0.028 \\[-0.05in]
     &     & Normal	& 0.98	& 0.44 	& 0.013 \\[-0.05in]
     &     & Inverted & 0.88	& 0.29	& 0.024 \\[-0.05in]
BMSV & \cite{BMSV}	& Inverted & 	&	& $\geq 0.01$ \\[-0.05in]
BKOT & \cite{BKOT}	& Normal	& 0.98	& 0.28	& 0.0001 - 0.0006 \\[-0.05in]
BO & \cite{BO}	& Normal	& 0.98 - 1.0 & 0.29 - 0.46 & 0.0014 \\[-0.05in]
BN & \cite{BN}	& Normal	& 1.0	& 0.36-0.39 & 0.0009 - 0.016\\[-0.05in]
BeMa & \cite{BeMa}	& Normal	& 0.93	& 0.40	& 0.012 \\[-0.05in]
BRT & \cite{BRT}	& Normal	& 0.99	& 0.35	& 0.0024 \\[-0.05in]
BW & \cite{BW}	& Normal	&	&	& O(0.01) \\[-0.05in]
CM & \cite{CM}	& Normal	& 1.0	& 0.41	& 0.014 \\[-0.05in]
DR & \cite{DR}	& Normal	& 0.98	& 0.40	& 0.0025 \\[-0.05in]
DMM & \cite{DMM}	& Normal	& 	&	& 0.0036 - 0.012 \\[-0.05in]
FO & \cite{FO}	& Normal	& 0.90	& 0.31 	& 0.04 \\[-0.05in]
GMN2 & \cite{GMN2}	& Normal	& $\leq 0.91$ & $\geq 0.52$ & 0.026 \\[-0.05in]
KR & \cite{KR}	& Normal	& 0.93	& 0.44	& 0.058 \\[-0.05in]
O & \cite{O}	& Normal	& 0.94	& 0.46 	& 0.0007 \\[-0.05in]
Ra & \cite{Ra}	& Normal	& 	&	& O(0.01) \\[-0.05in]
Ro & \cite{Ro}	& Normal	&	&	& 0.0056 \\[-0.05in]
      &    & Inverted &	&	& 0.036 \\[-0.05in]
ST & \cite{ST}	& Normal	& 0.99	& 0.46	& 0.0001 - 0.04 \\[-0.05in]
SP & \cite{SP}	& Normal	& 0.99	& 0.42 	& 0.0002 \\[-0.05in]
VR & \cite{VR}	& Normal	& 0.99 - 1.0 & 0.40 - 0.61 & 0.024 \\[-0.05in]
YW & \cite{YW}	& Normal	& 0.96	& 0.40	& 0.04 \\
\hline 
\end{tabular}
\end{table}
\clearpage

\begin{table}
\caption{\label{TableIV} Mixing Angles for $\mathbf{SO(10)}$ Models with 
  Lopsided Mass Matrices.}
\vspace*{0.1in}
\begin{tabular}{lr|c|c|c|c}
\multicolumn{2}{l}{\large Reference\ } & {\large \  Hierarchy \ } & 
  {\large \  $\sin^2 2\theta_{23}$ \ } & {\large \  $\tan^2 
  \theta_{12}$ \ } & {\large \  $\sin^2 \theta_{13}$ \ }\\[0.1in]
\hline \hline
A & \cite{A} & Normal & 0.98 - 1.0 & 0.38 - 0.50 & 0.002 - 0.003 \\[-0.05in]
AB & \cite{AB}	& Normal & 0.99	& 0.49	& 0.0002 \\[-0.05in]
BB & \cite{BB}	& Normal	& 0.97	& 0.40	& 0.0016 - 0.0025 \\[-0.05in]
JLM & \cite{JLM}	& Normal	& 1.0	& 0.41	& 0.019 \\[-0.05in]
Mae & \cite{Mae}	& Normal 	& 	&	& 0.048 \\[-0.05in]
P & \cite{P}	& Normal	& 0.99	& 0.17 - 0.29 & 0.0004 - 0.0025 \\
\hline
\end{tabular}
\end{table}
\clearpage
\newpage
\newpage

\begin{figure}
\includegraphics[scale=0.42]{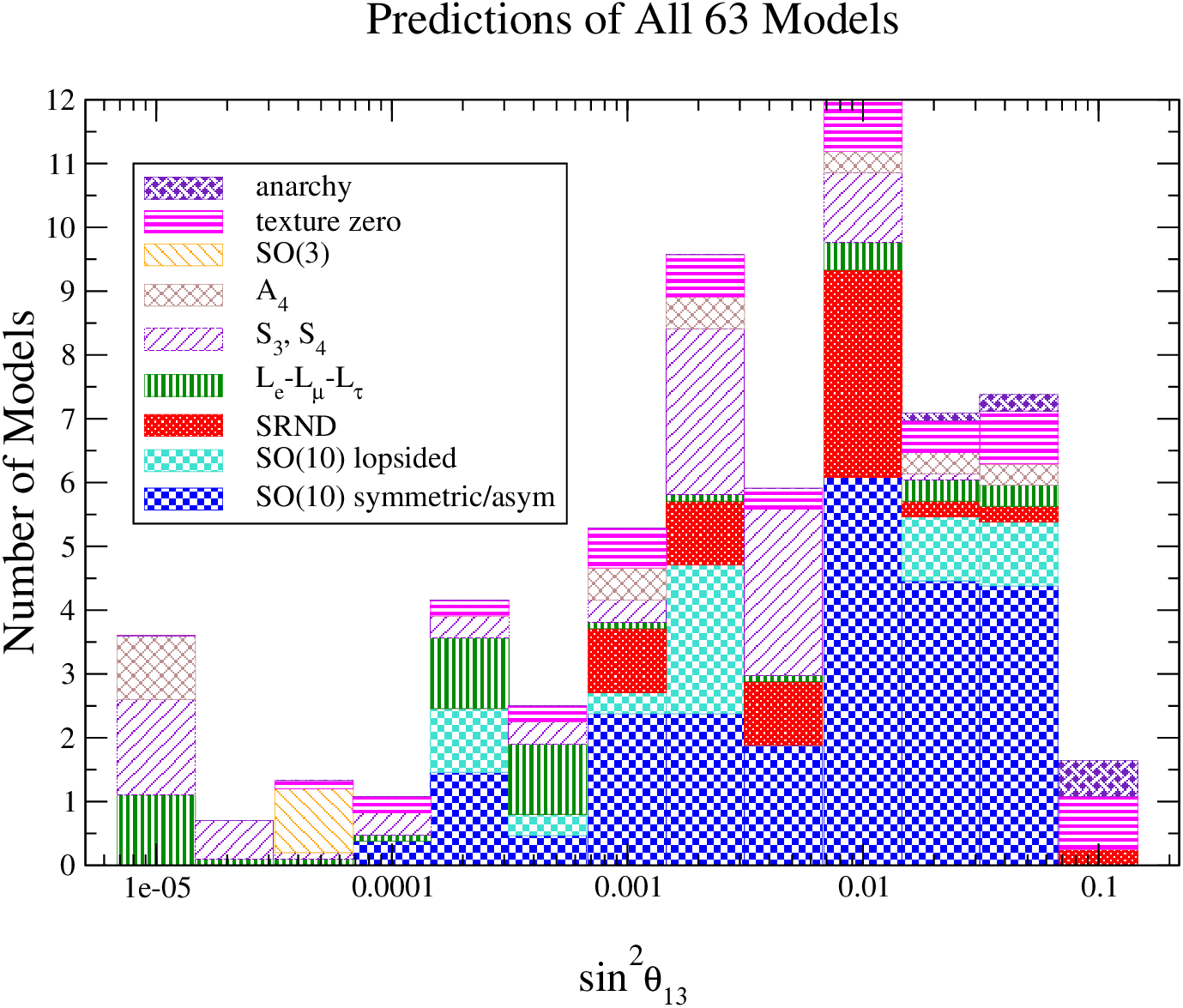}
\caption{\label{fig:1} Histogram of the number of models for each 
$\sin^{2}\theta_{13}$ including all 63 models.}
\end{figure}

\begin{figure}
\includegraphics[scale=0.42]{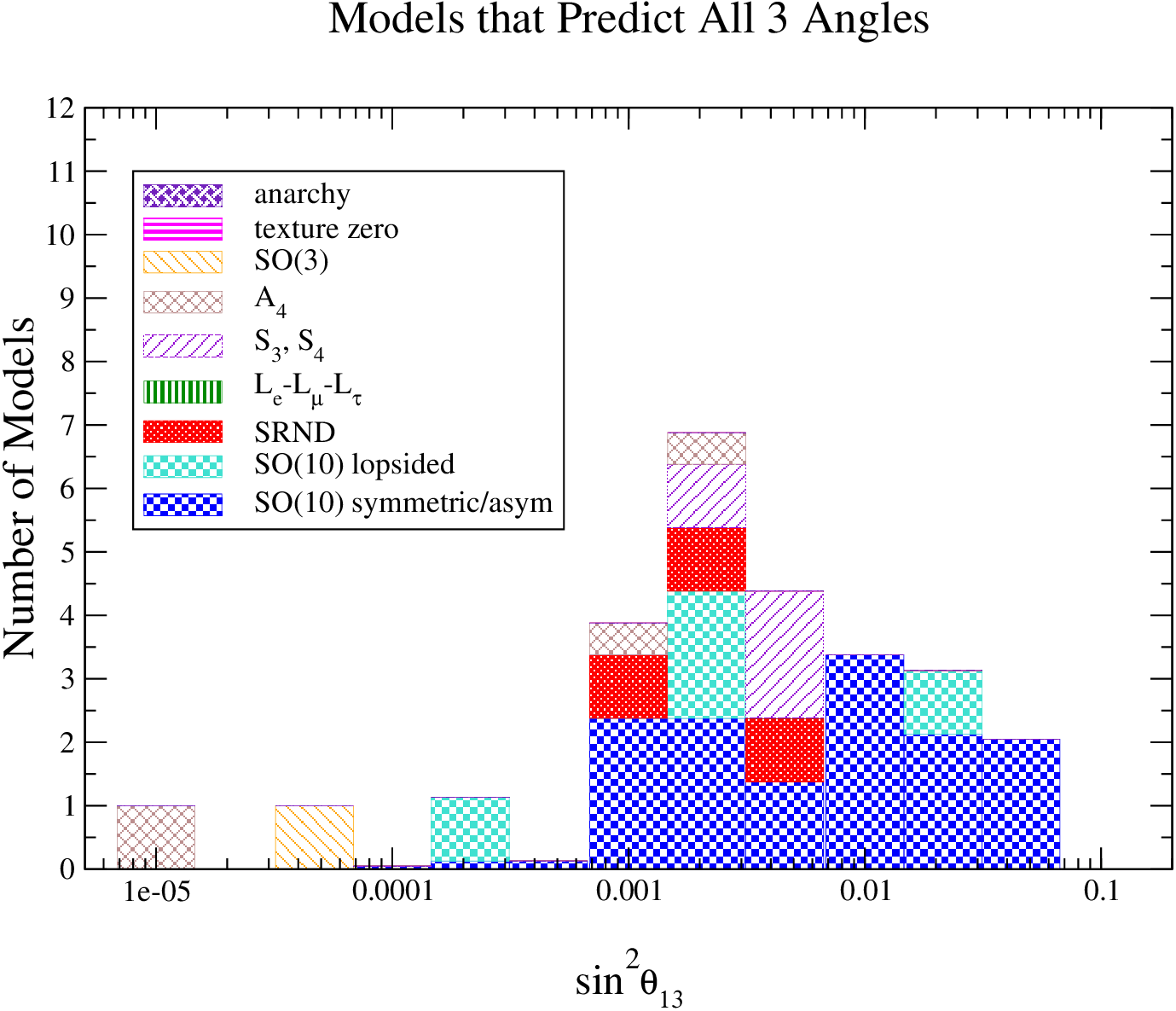}
\caption{\label{fig:2} Histogram of the number of models for each 
$\sin^{2}\theta_{13}$ that give accurate predictions for all three 
leptonic mixing angles.}
\end{figure}

\begin{figure}
\includegraphics[scale=0.42]{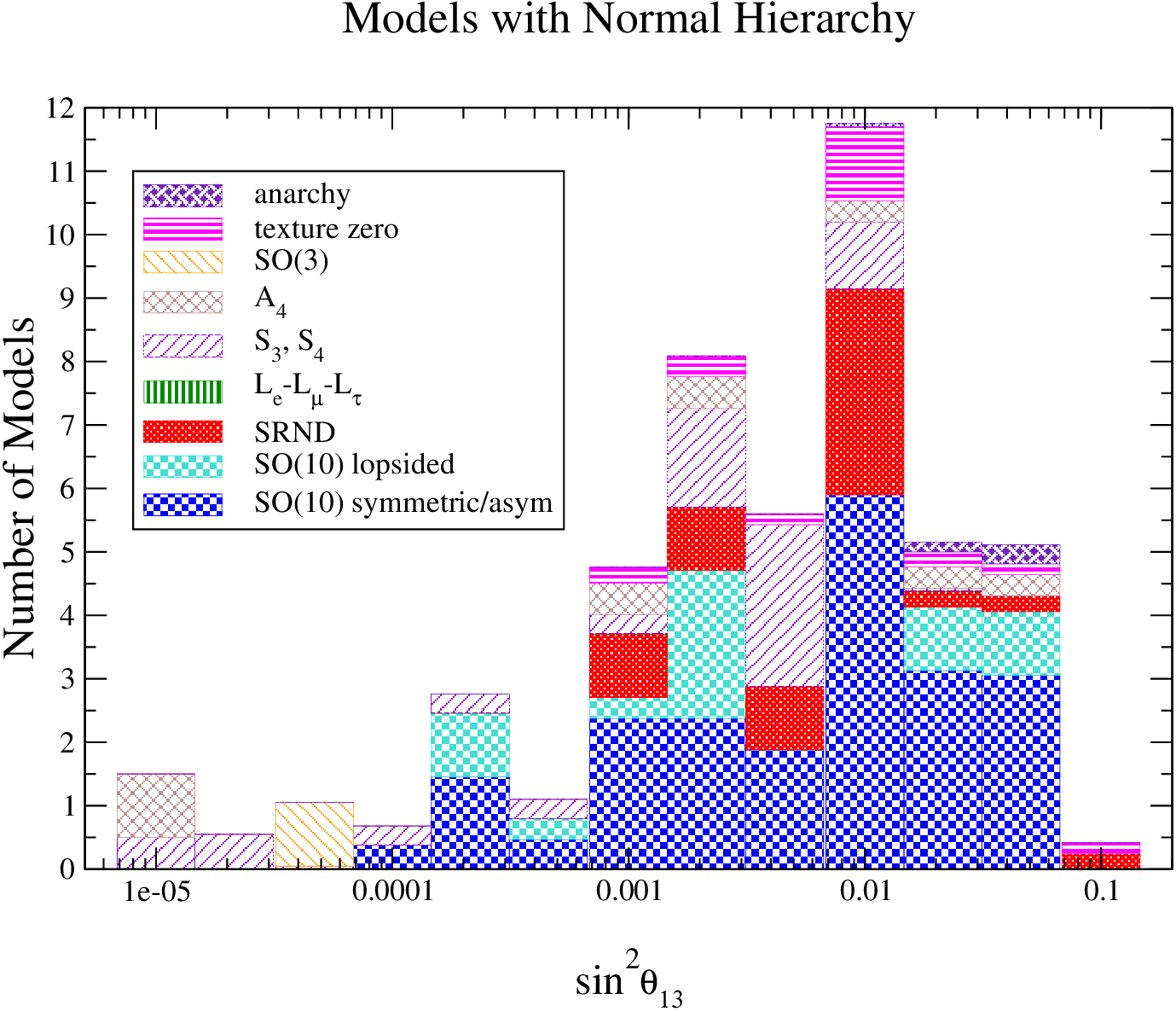}
\vskip 0.5in
\includegraphics[scale=0.42]{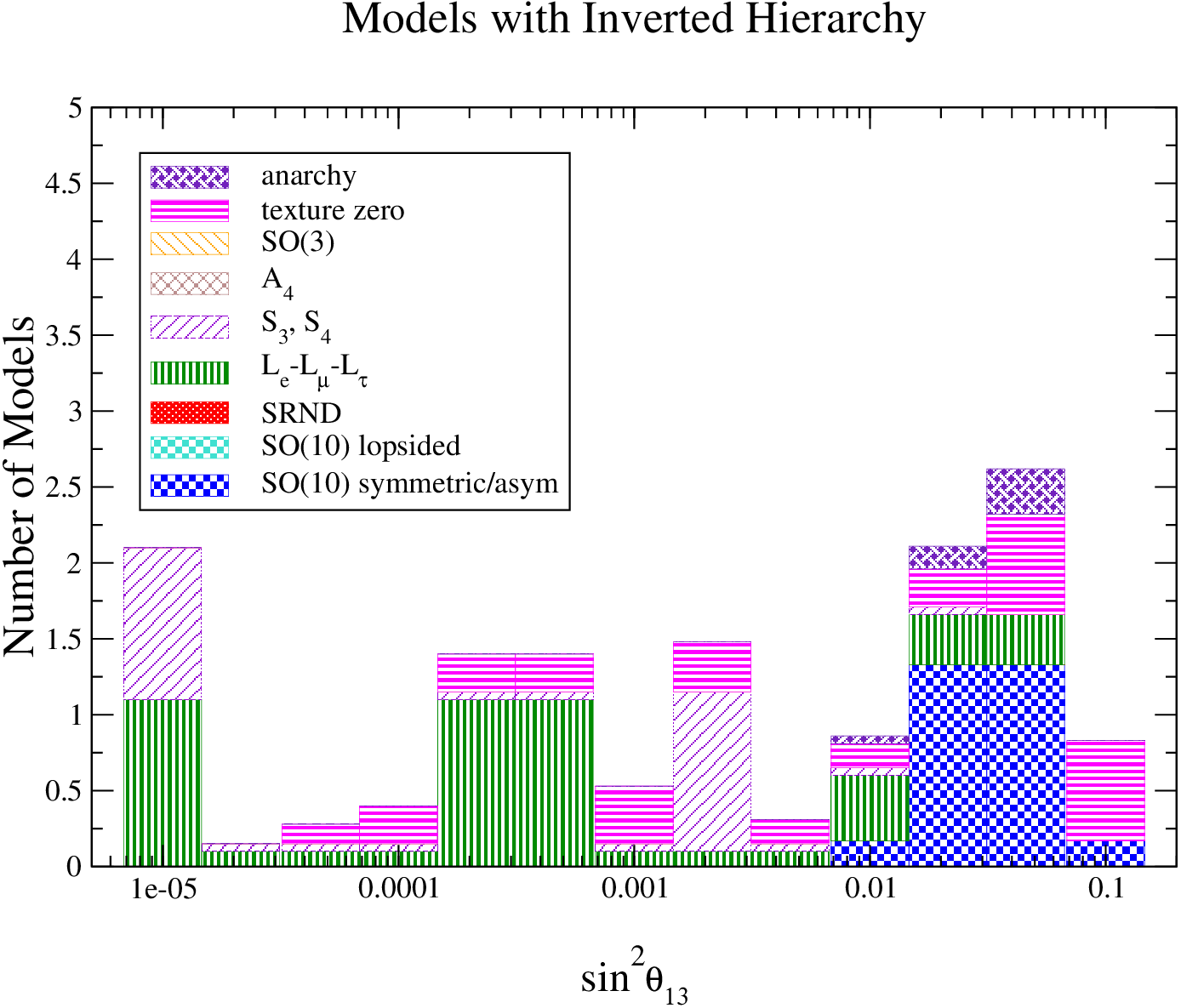}
\caption{\label{fig:3} Histograms of the number of models for each 
$\sin^{2}\theta_{13}$ where the upper diagram includes models that predict 
normal mass hierarchy, while the lower diagram includes models that 
predict inverted mass hierarchy.}
\end{figure}

\end{document}